\documentclass[12pt]{article}

\usepackage[utf8]{inputenc}
\usepackage[T1]{fontenc}
\usepackage[margin=1in]{geometry}
\usepackage{setspace}
\usepackage{amsmath}
\usepackage{booktabs}
\usepackage{threeparttable}
\usepackage{graphicx}
\usepackage{array}
\usepackage{pgfplots}
\pgfplotsset{compat=1.18}
\usepackage[authoryear,round]{natbib}
\usepackage{xurl}
\usepackage[colorlinks=true,linkcolor=blue,citecolor=blue,urlcolor=blue]{hyperref}
\title{\Large\textbf{Artificial Intelligence: Supply-Chain Chokepoints\\
and the Reach of Industrial Policy}\thanks{This project has a companion public dashboard at \url{https://aistackmap.org}, where all the series used in the paper, its underlying code, and its sources can be inspected. All errors are my own.}}

\author{Piyush Akimitsu\\ \textit{State University of New York, Department of Economics}\\ \small \href{mailto:econpi@icloud.com}{econpi@icloud.com}}

\date{\today}

\begin{document}
\maketitle

\begin{abstract}
\noindent Artificial intelligence depends on a stack of inputs, models on compute, compute on chips, and chips on electricity and refined minerals. This paper measures the concentration of each layer on one scale, the Herfindahl-Hirschman Index (HHI), from cited and reproducible data. Three findings follow. First, concentration forms a clear gradient. It is modest downstream, where public and regulatory attention is heaviest and model usage and cloud fall below the 1{,}800 mark United States agencies treat as highly concentrated. It rises steeply upstream, where public discourse is sparse. There advanced packaging scores 8{,}100, leading-edge lithography reaches the ceiling of 10{,}000, and one country dominates the production or refining of several critical minerals, gallium near that maximum. Second, these upstream layers are chokepoints and a strategic vulnerability for the whole AI economy. Antitrust cannot reach them, because they lie in foreign or state hands. Contesting them falls to export controls and domestic industrial policy. Third, placing reserves beside refining shows the concentration is built rather than geological, a variable industrial policy can move. A rise in an upstream input's price barely changes the cost of the product built from it, because the input is only a small share of that cost. A chokepoint's threat is therefore the loss of the input itself rather than a higher price.
\end{abstract}


\newpage
\doublespacing

\section{Introduction}
The public discussion about competition in artificial intelligence is mostly centered around the least concentrated layer. Most accounts ask whether a small number of firms at the downstream end of the stack, the developers of frontier models, are consolidating into an oligopoly. Measured on the index that antitrust authorities actually use, the Herfindahl-Hirschman Index, the downstream end is the least concentrated part of the stack. Frontier-model usage is spread across a dozen providers, which puts its index near 1{,}150. The cloud platforms that host most large training runs score 1{,}410. Both fall below the 1{,}800 mark at which United States agencies now treat a market as highly concentrated \citep{dojftc2023}.

One layer upstream, the scenario is different. A single firm performs close to ninety percent of the advanced packaging that current AI accelerators require. The machines that print leading-edge logic come from a single company. Three firms supply essentially all of the high-bandwidth memory. And upstream of the chips, one country produces almost all of the world's gallium. On the same index, these markets score above 8{,}000. Gallium approaches the theoretical ceiling of 10{,}000. The concentration the competition debate is searching for is real. However, it is simply located in three layers upstream of where the debate is focused.

This paper measures concentration across all layers of the AI input stack on one comparable scale and documents a clear gradient. Concentration is low downstream, where public and regulatory scrutiny is heaviest. It rises steeply toward the physical inputs upstream, where such scrutiny is thin and where a single firm or a single state often dominates. The paper's primary contribution is the measurement. The policy recommendations that follow rest on it.

\textbf{The organizing idea.} Artificial intelligence is produced from a stack of input layers, each a distinct market with its own concentration. For instance, a trained model is of little use without compute. Compute runs on chips, which are made of logic, memory, and packaging. Those components are in turn built with a short list of fabrication tools and process materials. The chips and the data centers that hold them run on electricity. Upstream of that electricity are mined and refined minerals. This is true even for renewable energy. Solar cells are built from refined silicon, wind turbines from rare-earth magnets and copper, and grid storage from graphite and lithium, so a shift toward renewables deepens the dependence on the upstream minerals rather than escaping it. This is the production-function view of capability, in the tradition that treats what an agent values as something manufactured from inputs rather than bought directly \citep{becker1965}. It carries a simple corollary that motivates the paper. Concentration in the upstream layers of the stack is convertible into leverage over everything downstream of it, because control of an input that no one else can supply is control of the output that requires it \citep{reytirole2007}.

\textbf{Four contributions.} First, the paper measures concentration across all layers of the stack on one comparable scale. Concentration in cloud, chips, packaging, lithography, memory, electricity, and a set of critical minerals is usually reported in different units by different communities, which makes the layers impossible to compare. The Herfindahl-Hirschman Index, which also expresses concentration as an equivalent number of equally sized firms \citep{adelman1969}, gives all of them a common scale and a common reference point in the merger guidelines. Second, all figures in the paper are cited, validated observations, each with a stated confidence tier. The full pipeline that produces the figures is public \citep{aistackmap}. The aim is that any number can be traced to its source and reproduced, which is not the norm in a literature that relies on proprietary market-research estimates. Third, it documents the gradient and draws out its uncomfortable implications. The most concentrated layers of the AI economy are largely beyond the reach of any antitrust authority, because they are located in foreign jurisdictions or in state hands. That is why they are contested with export controls and industrial policy rather than merger review. Fourth, it asks whether that concentration is geological or chosen, by placing reserves on the same index as production and refining. For all minerals with reported reserves, concentration is lowest in the ground and highest at the refinery. Gallium and silicon metal have no reported reserves at all, not because they are scarce but because they are ubiquitous. Gallium is a trace byproduct of aluminum ores with no deposit of its own. Silicon is smelted from quartz, one of the commonest minerals in the crust. In both cases the concentration is built at the processing stage, not inherited from geology. That makes it a question of capacity, which policy can move, not of endowment, which it cannot.

Three strands of literature bear on this question. None of them does what the paper does. The industrial-organization literature has documented a broad rise in firm concentration and markups over four decades \citep{deloecker2020,autor2020,grullon2019}, but it studies the economy at large rather than the input stack behind a single general-purpose technology. A newer literature on the economics and governance of artificial intelligence has concentrated its attention on the model and foundation-model layer \citep{bommasani2021,korinekvipra2025,cma2023}, which is precisely the layer this paper finds least concentrated. And the evidence on chips, minerals, and power is largely the province of technical and policy reports \citep{khan2021,iea2025cmo,usgs2025}, which measure each layer in its own units and rarely frame the result as concentration economics. The contribution here is to measure the concentration of these layers on one common index. The gradient is visible only once they are comparable.

\textbf{The gradient.} Figure~\ref{fig:gradient} presents it. Reading from the upstream end, gallium production scores close to 9{,}800, graphite processing 8{,}649, rare-earth processing 8{,}100, and cobalt refining 6{,}241, all dominated by one country. Advanced chip packaging scores 8{,}100 and leading-edge lithography reaches the ceiling of the index. High-bandwidth memory, a three-firm market, scores 4{,}174. The cloud layer is 1{,}410, and frontier-model usage is near 1{,}150. For comparison, United States agencies treat a market as highly concentrated above 1{,}800, and several input layers score three to five times that. Concentration in the AI economy is not absent from the downstream end of the stack. It is modest there and severe upstream. It is most severe where competition policy has the least reach.

The rest of the paper proceeds as follows. Section \ref{sec:lit} places the paper among four strands of literature it draws on and departs from. Section \ref{sec:framework} sets out the production-function framework, the index, and how an input's cost share decides whether a chokepoint threatens through price or through lost supply. Section \ref{sec:layers} measures the layers one by one, from the downstream models to the upstream minerals, and finds concentration rising at each step upstream. Section \ref{sec:gradient} draws the layers together and locates the point at which antitrust reaches its limit and statecraft begins. Section \ref{sec:geology} asks whether the concentration is geological or chosen. Comparing reserves with production and refining, it finds the concentration largely built. It then turns to the industrial-policy literature for what can rebuild capacity. Section \ref{sec:limits} discusses limitations and extensions. Section \ref{sec:conclusion} concludes.

\section{Related Literature}\label{sec:lit}
This paper draws on four strands of literature, the measurement of market concentration, the economics of production structure and supply networks, the political economy of concentrated power, and the nascent research on concentration in artificial intelligence. Each contributes a tool or a fact. Each leaves a gap the paper fills.

\subsection{The measurement of concentration and its limits}
The index at the center of the paper has a long history. \citet{hirschman1964} records that he introduced the sum of squared market shares in 1945, before Herfindahl, which is why the measure carries both names. \citet{adelman1969} supplied the interpretation that makes it useful across very different markets. The reciprocal of the index is a numbers-equivalent, the number of equal-sized firms that would produce the same value. An index of 2{,}500 is the concentration of four equal firms and an index near 10{,}000 is the concentration of one. The index is embedded in antitrust practice through the merger guidelines, which set the thresholds at which a market counts as concentrated and a merger draws antitrust scrutiny \citep{dojftc2010,dojftc2023}. \citet{tirole1988} remains the standard treatment of the underlying theory of market power.

A large empirical literature has used these and related measures to argue that the American economy grew more concentrated over the past forty years. \citet{deloecker2020} estimate that average markups rose from twenty-one percent over marginal cost in 1980 to sixty-one percent, with the increase concentrated in the upper tail of firms. \citet{autor2020} document rising product-market concentration across most sectors since the early 1980s and tie it to the decline in labor's share of income, through the reallocation of activity toward a few high-markup superstar firms. \citet{grullon2019} find that more than three-quarters of United States industries became more concentrated after the late 1990s, with higher margins and no accompanying efficiency gains. \citet{barkai2020} shows that labor and capital shares fell together while pure profits rose, a pattern that points to markups rather than to a simple substitution between factors. \citet{philippon2019} assembles these threads into the argument that American markets became less competitive as enforcement weakened, using European markets as a counterfactual.

This literature has also produced a set of methodological cautions. They bear directly on any exercise that measures concentration. Two 2019 symposia in the \textit{Journal of Economic Perspectives}, on markups and on antitrust, set them out. \citet{berry2019} warn against reviving the discredited structure-conduct-performance inference that reads market power directly off concentration, and insist on industry-specific analysis. \citet{syverson2019} stresses that conclusions depend on how market power is defined and measured, and that accounting-based markup estimates rest on strong assumptions. \citet{shapiro2019} argues that concentration is not by itself a reliable measure of market power. A broad or national figure need not correspond to any relevant antitrust market. Even within a well-defined market, higher concentration can reflect efficient firms winning share rather than weaker competition. What matters is why concentration rose. That can be judged only market by market. Two recent papers show how much the verdict depends on the boundary of the market. \citet{rossihansberg2021} find that national product-market concentration rose from 1990 to 2014 while local concentration fell over the same period. \citet{benkard2026} find that in narrow, antitrust-style product markets the median index actually declined between 1994 and 2019.

A concentration figure is informative when the market is drawn at the boundary where substitution is hard, and misleading when the boundary is too broad. The layers this paper measures are drawn at that boundary. Each is a single step in building an AI system, defined by a supply-chain stage rather than by a broad industry aggregate. Advanced packaging, extreme-ultraviolet lithography, high-bandwidth memory, and gallium refining each have few or no ready substitutes. The general concentration debate is about how broadly to draw the market. For these layers that question is already settled, because each is a single step that a user cannot substitute around. A high index for one of them therefore reflects a genuine bottleneck, not an artifact of a market drawn too broadly.

An AI system's capability is an output produced from inputs, which makes the economics of production structure the natural framework. \citet{becker1965} is the antecedent for treating what is ultimately valued as something produced, in his case by households combining time and market goods. A similar logic applies to a firm assembling a model from compute, data, chips, and power. The structure of the chain that supplies those inputs then determines where market power can reside. The vocabulary for that structure is worth fixing precisely, because the paper relies on it throughout. A firm is upstream when it supplies an input to another firm, and downstream when it buys that input and sells nearer the final user. A supply chain runs from the raw material at its most upstream point to the finished good at its most downstream one \citep{tirole1988}. \citet{spengler1950} showed that a chain of successive monopolies, each an upstream seller to the next, produces less and prices higher than a single integrated firm. This result is known as double marginalization. It is the first formal reason to care about concentration at each separate layer rather than only in the final market. \citet{reytirole2007} generalize this to foreclosure, the use of control over a bottleneck input to deny downstream rivals access to it, which is the mechanism by which a concentrated upstream layer exercises power over the downstream layers that draw on it.

A growing body of research on production networks examines how shocks and power travel through such chains. \citet{acemoglu2012} show that when production is linked through input-output relationships, idiosyncratic shocks to individual suppliers need not cancel out but can generate aggregate fluctuations, with the network's structure governing how far and how fast disturbances spread. \citet{carvalho2019} survey the field. The empirical work makes this concrete. \citet{barrot2016} find that natural-disaster shocks to suppliers impose large losses on their customers, and that the damage is concentrated precisely when the input is specific and hard to substitute. \citet{carvalho2021} trace how the 2011 Japanese earthquake propagated upstream and downstream through supplier links into a sizeable aggregate effect. \citet{baietal} document the same for the pandemic era, tracing the 2021 inflation surge largely to disruptions at the shipping chokepoints of the global supply chain. The message for this paper is that a specific, hard-to-substitute input controlled by one supplier is not only a source of static market power but a channel through which a single failure or a single decision propagates across everything downstream of it. \citet{antras2013} and \citet{baldwinfreeman2022} add that the organization and the fragility of these chains depend systematically on position within them, and that measured exposure to foreign shocks understates true exposure because upstream linkages are hidden and easily overlooked.

Two recent papers connect this structure directly to policy. They are the closest antecedents to this paper. Technology does not leave the shape of a supply chain fixed. In related work on healthcare price regulation, I find that the effects of price and cost controls are mediated by technology and by non-price, quality competition, so that the same regulation restructures a market differently depending on the local availability of the enabling technology \citep{akimitsu2024}. The general lesson is that the impact of a regulation is settled by conditions in the supply of the inputs it touches rather than in the regulated market alone. This lesson carries directly to the AI stack, where the technologies of refining, chipmaking, and artificial intelligence itself reshape the elasticity of the inputs they draw on. \citet{mulligan2024} reaches a complementary conclusion from the direction of price controls, showing that compliance is achieved by altering the mix of production factors along the supply chain, so that the incidence of a regulation is determined by conditions upstream. Both results imply that the layers where substitution is hardest, the concentrated upstream layers of the AI stack, are where the economics of both market power and regulation are ultimately settled. That is a reason to measure them, not to assume them away.

\subsection{Economic statecraft and the political economy of concentration}
When a concentrated input is located in another country, its economics becomes geopolitics. \citet{hirschman1945} established the founding idea, that asymmetric trade dependence is convertible into political power, as a large state steers commerce toward smaller partners and then exploits the dependence it has created. \citet{baldwin1985} built the framework for evaluating such economic tools of statecraft. \citet{farrell2019} gave the contemporary version its name. States with jurisdiction over the central nodes of global economic networks can weaponize them, through a chokepoint effect that denies adversaries access and a panopticon effect that gathers information from the flows they host. \citet{clayton2023} formalize the economics of this leverage, modeling how a dominant state coordinates threats across economic relationships to extract concessions. This is the vocabulary for the upstream end of the AI stack, where the relevant unit is not a firm with a large market share but a country with a near-monopoly over refining, and where the exercise of power takes the form of an export license rather than a price.

Concentration also feeds back into domestic politics. \citet{zingales2017} argues that market power gives a firm both the means and the motive to acquire political power, producing a self-reinforcing loop in which economic and political concentration reinforce each other. \citet{rajanzingales2003} develop the related claim that the chief threat to open markets is capture by entrenched incumbents. \citet{acemoglurobinson2008} show formally how elites can invest in de facto power to blunt changes in formal institutions. These results are the reason a paper about input concentration is not only a paper about prices. Concentrated control of the inputs to a general-purpose technology is a durable form of economic power. Durable economic power tends to entrench itself.

The strand of literature closest to this paper is also the one it most directly answers. Scholarly and regulatory attention to concentration in artificial intelligence has settled, understandably, on the layer that is most visible, the models. \citet{bommasani2021} named the category foundation models and organized a large research agenda around it. The naming itself marks where the field's attention crystallized. The most rigorous economics in this vein, \citet{korinekvipra2025}, shows that the economies of scale and scope in compute, data, and talent give the foundation-model market a structural tendency toward concentration. National competition authorities have focused on the same layer. The United Kingdom's flagship review is scoped explicitly to foundation models, with compute entering only as a barrier to entry \citep{cma2023}. The United States has only recently begun to examine the layer just upstream of the models, in its study of cloud-and-model partnerships and their effect on access to inputs such as computing power \citep{ftc2025}.

A smaller body of work has begun to argue that the layers upstream of the models are the concentrated ones. \citet{sastry2024} observe that compute is uniquely governable because it is detectable, excludable, quantifiable, and produced through an extremely concentrated supply chain. \citet{vipra2023} describe that supply chain as profoundly monopolized at key points by one or a few firms. \citet{widder2024} show that the language of open models obscures the concentration in the compute and infrastructure upstream of them. The facts of the hardware chokepoints are well documented in research on semiconductors and energy. The chip supply chain is geographically concentrated, with single-supplier nodes at lithography and design software \citep{khan2021,miller2022,bown2020}. One country dominates the mining, and more sharply the refining, of critical minerals \citep{iea2024cmo,iea2025cmo,usgs2025,nassar2020}. The electricity demand of AI data centers is colliding with a grid that cannot connect new supply quickly \citep{masanet2020,ieaenergyai2025,shehabi2024,rand2024}.

The gap this paper fills is left open by all three areas of research. Research on economy-wide concentration does not examine the AI stack. Research on concentration in AI examines mostly the downstream end, the models. Evidence of concentration in the upstream layers exists, but it is fragmented. Different research communities measure it in different units. None places it on a common scale. Without a common scale, the gradient across the stack cannot be seen. This paper measures concentration in all layers on one comparable index, with public and reproducible provenance. It finds that concentration rises as one moves upstream. None of the three areas of research could establish this gradient on its own. The rest of the paper builds on it.

\section{A Framework for Measuring Concentration Across the Stack}\label{sec:framework}
\subsection{The stack as a production function}

AI capability is treated as an output produced from a stack of inputs, in the production-function tradition of \citet{becker1965}. A deployed model is produced from a trained model and the compute that serves it. The trained model is produced from compute, data, and algorithms. Compute is produced from chips. A chip is produced from logic, memory, and packaging, using design software, lithography, and fabrication tools. The chips and the data centers that house them run on electricity. Electricity itself depends on mined and refined minerals. Each step in this chain is a layer, a market with its own structure. The paper measures the concentration of all layers on one index.

The chain runs in one direction. The vertical image of a stack, however, leaves that direction ambiguous. The paper fixes it. The deployed model is the downstream end, the good the final user buys. The mined mineral is the upstream end, the raw input on which everything else is built. Reading down the stack, from models toward minerals, is therefore reading upstream, toward the source. The paper's central finding is that concentration rises upstream, toward the raw inputs. The paper sorts the layers into three bands. A band is a group of adjacent layers. The downstream band holds two layers, the model and the cloud that serves it. The midstream band holds the layers that make a chip, design software, memory, packaging, lithography, and fabrication. The upstream band holds the physical layers, electricity and the mined and refined minerals. Where a finer position matters, the paper names both the band and the place within it. Electricity is the upstream-top layer, nearest the chips. Mining is the upstream-bottom layer, at the base of the stack. Upstream, midstream, and downstream can describe two different scales. One convention keeps them apart. A single mineral is itself a supply chain, with its own upstream, midstream, and downstream, running from mining to refining to finished material. A stage inside a single mineral is therefore named directly, as mining or refining. The three whole-stack bands keep the terms upstream, midstream, and downstream.

Gallium runs through the paper as a touchstone. It is a metal refined almost entirely in one country, making up a negligible share of the cost of the chip that needs it, and lacking an ore body of its own. Each property returns below.

\subsection{Cost share and the price channel}
The index measures how concentrated a layer is. It does not by itself say how much that concentration can raise the cost of the products built on it. That depends on a second quantity, the share of a product's cost that the input accounts for. Write $C$ for the unit cost of a product and $\theta_i$ for the share of that cost taken by input $i$. Then $\theta_i = p_i x_i / C$, with $p_i$ the input's price and $x_i$ the quantity used per unit of output. The elasticity of unit cost with respect to the input's price equals that cost share, a standard implication of cost minimization,
\begin{equation}
\frac{\partial \ln C}{\partial \ln p_i} = \theta_i . \label{eq:passthrough}
\end{equation}
A price rise in a concentrated input passes through to the product in proportion to $\theta_i$. Two cases follow. When the input is a large share of cost, the price rise reaches the product in full. When the share is small, as it is for gallium in a chip, even a large price rise barely moves the cost. As $\theta_i$ approaches zero the price channel closes. What remains is the quantity. If $x_i$ cannot be obtained, the product cannot be made at any price. The exposure then runs through the availability of the input rather than its price. Section \ref{sec:geology} shows that the mineral inputs to AI hardware are of this second kind. Their concentration threatens the stack through availability rather than through price.

\subsection{The index and its conventions}
For a market with participant shares $s_i$, the Herfindahl-Hirschman Index is
\begin{equation}
\text{HHI} = 10{,}000 \times \sum_i s_i^2 , \label{eq:hhi}
\end{equation}
bounded above by 10{,}000 when one participant holds the entire market, and interpretable through its reciprocal as an effective number of equal-sized participants \citep{adelman1969}. Gallium, produced almost entirely in one country, scores near this ceiling. I anchor all values to the same public reference point. The United States merger guidelines treat a market above 1{,}800 as highly concentrated \citep{dojftc2023}, a threshold lowered from 2{,}500 in 2010 \citep{dojftc2010}.

Three conventions keep the reported values conservative, meaning each is a lower bound on true concentration rather than an overstatement. Each participant's share is its quantity as a fraction of the market total. The index squares these shares and adds them. Two choices therefore shape the result, the market total the shares are measured against and the treatment of participants a source does not report individually. The first convention concerns the residual, the small participants a source reports only as a combined total rather than individually. I square the share of each named participant. I do not square the residual. The residual is not ignored. It stays in the market total and still reduces all the named shares. What I leave out is only the residual's own squared term. Omitting that term is equivalent to assuming the residual is spread across many tiny participants. Any more concentrated split would only raise the index. The reported value is therefore a lower bound. The second convention concerns the market total itself. I use the figure the source publishes, the world or national total, rather than one built by summing the listed participants. Where the listed quantities add up to more than the published total, I use the larger of the two. No share can then exceed one. The index cannot exceed its ceiling of 10{,}000. The third convention concerns revision. The mineral series are assembled from successive annual editions of the USGS Mineral Commodity Summaries. When two editions report different figures for the same year, I use the later edition, which incorporates corrections.

\subsection{Confidence tiers and provenance}
Layers differ in how directly their concentration can be observed. The paper labels that difference rather than hiding it. Each figure carries one of four ordered confidence tiers. A value is \textit{measured} when it comes from a primary count or an official statistic, as with gallium's country shares from the USGS. It is \textit{derived} when it is computed from measured components, as with a mineral's production index. It is \textit{modeled} when it comes from a documented estimate with stated assumptions, as with cobalt refining shares from an industry body. It is \textit{proxy} when it is a substitute with a stated gap, as with cloud revenue share used in place of installed compute capacity. The tier stays attached to each figure. A reader can therefore weigh a proxy differently from a measured value.

\subsection{Data, replication, and vintage}
All figures in this paper come from a public pipeline. The ingest code, the metric definitions, and the download ledger are published alongside the data. Each observation is stored in a data panel with one row per measurement. Each row carries its bottleneck, metric, date, geography, value, unit, source identifier, and retrieval timestamp. Any number in the tables and figures can therefore be located, checked against its source, and recomputed \citep{aistackmap}. The panel is rebuilt from the per-source files rather than edited by hand.

The date of each figure matters, because the underlying series update at different frequencies. The mineral figures are annual and come from successive editions of the USGS Mineral Commodity Summaries. A later edition can revise a given year's value. When two editions disagree, the paper uses the later one, as stated above. The chip, cloud, model, electricity, and price series update continuously or quarterly. All values reported here are as of July 2026. Each observation's retrieval date is recorded in the public data. A reader comparing these figures with a later version of the data should expect the fast-moving series, model usage and the chip market-tracker shares in particular, to have changed.

\begin{figure}[htbp]
\centering
\begin{tikzpicture}
\begin{axis}[
  xbar,
  width=0.80\textwidth, height=0.55\textwidth,
  xmin=0, xmax=22,
  xlabel={Share of usage (percent)},
  symbolic y coords={Other,StepFun,Qwen,Moonshot,NVIDIA,MiniMax,Z-AI,OpenAI,Google,Anthropic,DeepSeek,Xiaomi,Tencent},
  ytick=data,
  nodes near coords, nodes near coords align={horizontal},
  every node near coord/.append style={font=\footnotesize},
  bar width=8pt, enlarge y limits=0.05,
  xmajorgrids, tick align=outside,
]
\addplot[fill=gray!55, draw=black!40] coordinates {
 (19.1,Tencent) (16.0,Xiaomi) (13.9,DeepSeek) (11.3,Anthropic) (6.9,Google)
 (6.8,OpenAI) (6.5,Z-AI) (5.8,MiniMax) (5.4,NVIDIA) (1.9,Moonshot)
 (1.9,Qwen) (1.2,StepFun) (3.4,Other)
};
\end{axis}
\end{tikzpicture}
\caption{Frontier-model usage by provider, as of 20 July 2026. No provider holds a fifth of the market. The usage-weighted index is 1{,}149 with the residual pooled. Chinese developers account for about 66 percent of the total. Source, OpenRouter \citep{openrouter2026}.}
\label{fig:usage}
\end{figure}
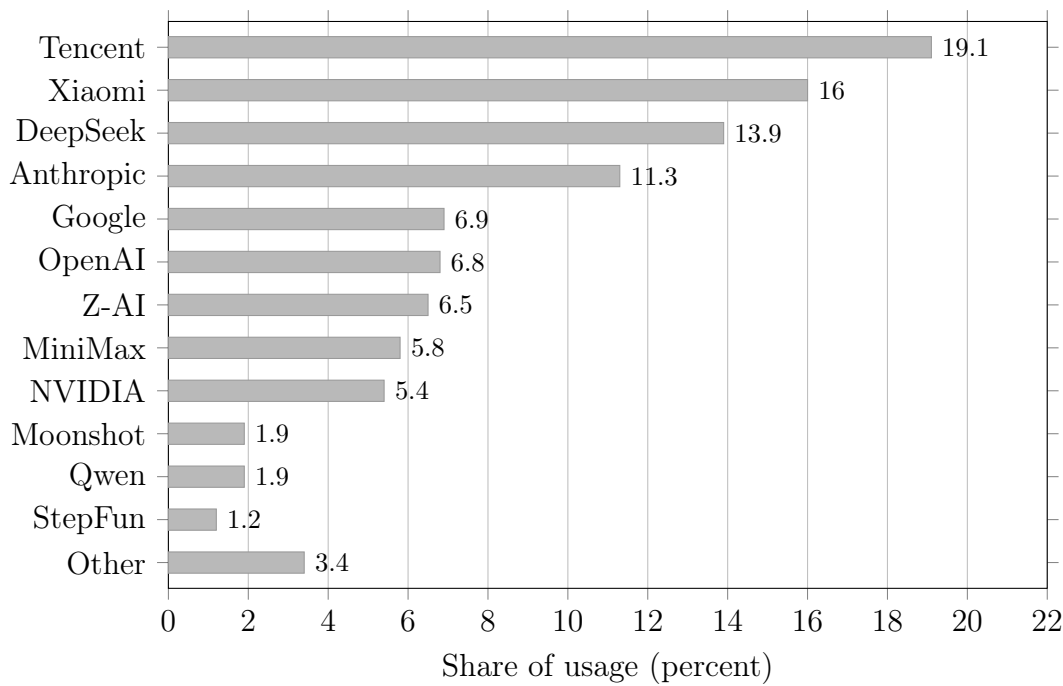

One index is computed in the text itself, the usage-weighted index for the frontier-model market. It comes from the provider shares in Figure \ref{fig:usage} through equation~\eqref{eq:hhi}, with the residual left unsquared as above, giving 1{,}149 for 20 July 2026. The other indices are either reported directly by their source or computed by the same pipeline. For each mineral, the paper reports the concentration index rather than the full country-by-country production composition behind it.

\section{The AI Stack, Layer by Layer}\label{sec:layers}
\begin{figure}[htbp]
\centering
\begin{tikzpicture}
\begin{axis}[
  xbar,
  width=0.88\textwidth, height=0.52\textwidth,
  xmin=0, xmax=10800,
  xlabel={Herfindahl-Hirschman Index},
  symbolic y coords={EUV lithography,Gallium production,Graphite processing,CoWoS packaging,Rare-earth processing,Cobalt refining,HBM memory,Cloud infrastructure,Frontier-model usage},
  ytick=data,
  nodes near coords, nodes near coords align={horizontal},
  every node near coord/.append style={font=\footnotesize},
  bar width=11pt,
  enlarge y limits=0.07,
  xmajorgrids, tick align=outside,
]
\addplot[fill=gray!55, draw=black!40] coordinates {
 (1149,Frontier-model usage)
 (1410,Cloud infrastructure)
 (4174,HBM memory)
 (6241,Cobalt refining)
 (8100,Rare-earth processing)
 (8100,CoWoS packaging)
 (8649,Graphite processing)
 (9804,Gallium production)
 (10000,EUV lithography)
};
\draw[dashed, thick] ({rel axis cs:0.1667,0}) -- ({rel axis cs:0.1667,1});
\node[anchor=south west, font=\scriptsize] at ({rel axis cs:0.1667,0.86}) {1{,}800 highly-concentrated line};
\end{axis}
\end{tikzpicture}
\caption{The concentration gradient. Frontier-model usage and cloud fall below the merger-guideline threshold. The chip and mineral layers upstream of them are three to five times more concentrated. Sources as in Table \ref{tab:hhi}.}
\label{fig:gradient}
\end{figure}
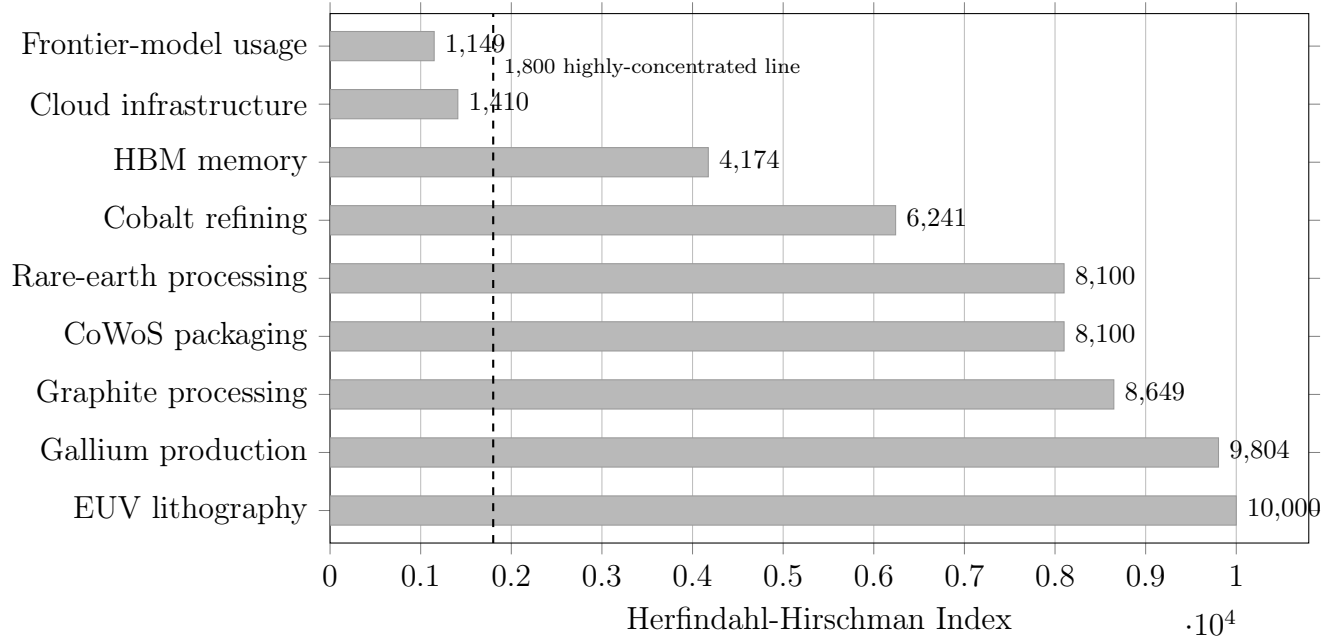

Table \ref{tab:hhi} and Figure \ref{fig:gradient} show the headline concentration of each layer on the common index. The rest of this section reads the stack from downstream to upstream, adding the finer detail within each layer.

\begin{table}[htbp]
\centering
\begin{threeparttable}
\caption{Concentration across the AI stack, on one index}
\label{tab:hhi}
\small
\begin{tabular}{@{}llcc@{}}
\toprule
Layer & Market or step & Leading share & HHI \\
\midrule
Models      & frontier-model usage\tnote{a}      & 19\% (top firm)  & $\approx$1{,}150 \\
Cloud       & hyperscale infrastructure\tnote{b} & 63\% (top 3)     & 1{,}410 \\
Memory      & high-bandwidth memory (HBM)\tnote{c} & $\approx$57\% (top firm) & 4{,}174 \\
Packaging   & advanced packaging (CoWoS)\tnote{d} & 90\% (TSMC)      & 8{,}100 \\
Lithography & EUV scanners\tnote{e}              & 100\% (ASML)     & $\approx$10{,}000 \\
Minerals    & cobalt refining\tnote{f}           & 79\% (China)     & 6{,}241 \\
Minerals    & rare-earth processing\tnote{f}     & 90\% (China)     & 8{,}100 \\
Minerals    & graphite processing\tnote{f}       & 93\% (China)     & 8{,}649 \\
Minerals    & gallium production\tnote{g}        & 99\% (China)     & 9{,}804 \\
Power       & data-center load, grid\tnote{h}    & \multicolumn{2}{c}{constraint, not an HHI} \\
\bottomrule
\end{tabular}
\begin{tablenotes}[flushleft]\footnotesize
\item All figures are compiled in the AI Stack Concentration Dashboard \citep{aistackmap}, where each is a validated, sourced observation with a confidence tier. Leading share is the top single firm or country, except where marked top~3. All values are as of July 2026.
\item[a] Author's calculation from OpenRouter provider usage shares, 20 July 2026, residual pooled \citep{openrouter2026}. A usage proxy for the model market.
\item[b] Synergy Research Group, cloud infrastructure services, 2025.
\item[c] Counterpoint Research, 2025. The three suppliers, SK Hynix, Samsung, and Micron, hold the entire market.
\item[d] TrendForce, 2025. TSMC's share of advanced (CoWoS) packaging.
\item[e] ASML is the sole maker of EUV scanners \citep{khan2021,miller2022}, so the index is at its ceiling by construction.
\item[f] Cobalt Institute (cobalt refining), IEA (rare-earth processing), and Benchmark Mineral Intelligence (graphite processing), 2024.
\item[g] USGS Mineral Commodity Summaries \citep{usgs2025}, primary low-purity gallium, 2025 estimate.
\item[h] Data centers reached 4.4\% of U.S.\ electricity in 2023 \citep{shehabi2024}. The binding constraint is a roughly five-year grid-interconnection queue \citep{rand2024}, not a market share.
\end{tablenotes}
\end{threeparttable}
\end{table}

\subsection{Models and cloud}
The downstream end of the stack is the least concentrated and the most contested among firms. It is the layer the AI-concentration literature examines most closely \citep{bommasani2021,korinekvipra2025,cma2023,ftc2025}. Measured by usage, the frontier-model market is fragmented. Provider shares on the OpenRouter routing platform spread across a dozen suppliers, the largest holding 19 percent and no other reaching 16, for a usage-weighted Herfindahl-Hirschman Index near 1{,}150 with the residual pooled \citep{openrouter2026}. The organizations that produce frontier models are, by one measure, less concentrated still, with a developer-level index around 490 in 2025 \citep{epoch2026}. The cloud infrastructure just upstream of the models is more concentrated but remains moderate. The three hyperscalers hold a combined 63 percent, a three-firm concentration ratio. The index itself is 1{,}410 (Synergy Research Group). All three figures fall below the 1{,}800 threshold.

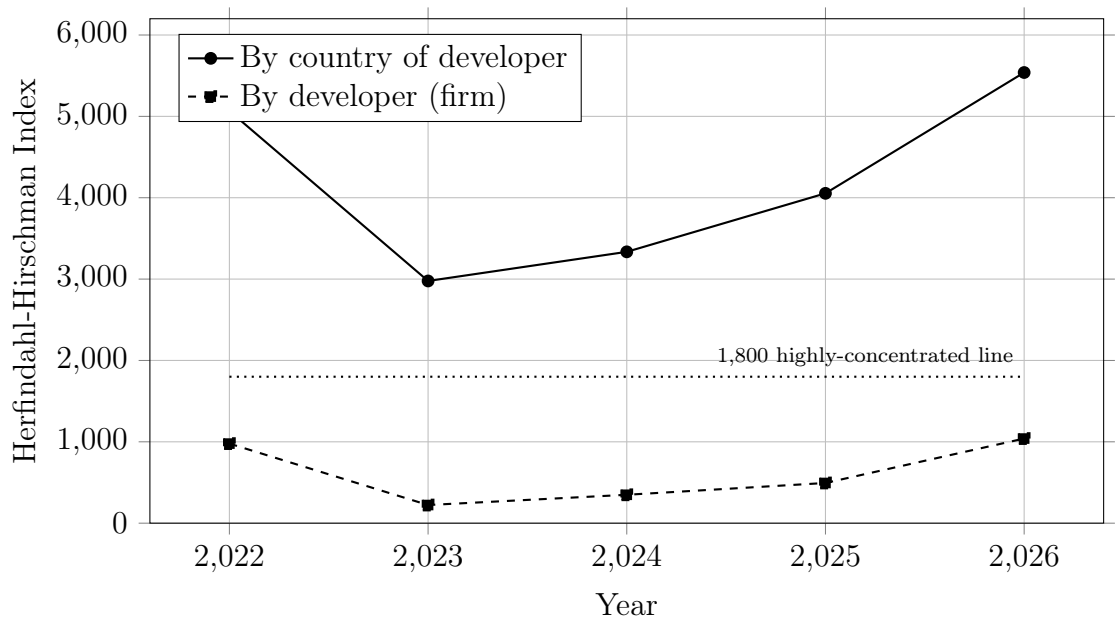
\begin{figure}[htbp]
\centering
\begin{tikzpicture}
\begin{axis}[
  width=0.86\textwidth, height=0.5\textwidth,
  xlabel={Year}, ylabel={Herfindahl-Hirschman Index},
  xtick={2022,2023,2024,2025,2026},
  ymin=0, ymax=6200,
  legend pos=north west, legend cell align={left},
  grid=major, tick align=outside,
]
\addplot[thick, mark=*] coordinates {(2022,5067) (2023,2977) (2024,3335) (2025,4054) (2026,5539)};
\addlegendentry{By country of developer}
\addplot[thick, dashed, mark=square*] coordinates {(2022,978) (2023,223) (2024,349) (2025,494) (2026,1040)};
\addlegendentry{By developer (firm)}
\draw[dotted, thick] (axis cs:2022,1800) -- (axis cs:2026,1800);
\node[anchor=south east, font=\scriptsize] at (axis cs:2026,1800) {1{,}800 highly-concentrated line};
\end{axis}
\end{tikzpicture}
\caption{The model layer, fragmented by firm but concentrated by country. The firm-level index stays near 500 while the country-of-developer index rises past 5{,}000, well above the 1{,}800 threshold. The counts behind each year are 30, 121, 168, 164, and 23 frontier models for 2022 through 2026. The 2026 figure is a partial year through July and is therefore provisional. Source, Epoch AI \citep{epoch2026}.}
\label{fig:models}
\end{figure}

One feature of the model layer complicates the simple reading and foreshadows the rest of the paper. The layer that is fragmented by firm is concentrated by country. Measured by the nationality of the organizations producing frontier models, the index rose from about 2{,}977 in 2023 to about 5{,}539 in 2026, well into the highly concentrated range, as capability consolidated in the United States and China (Figure \ref{fig:models}). The firm-level and country-level indices differ by an order of magnitude, the first staying near 500 while the second exceeds 5{,}000. The competitive picture and the geopolitical picture already diverge at the downstream end of the stack.

The two countries that account for almost all of that concentration are close to each other in output. The balance has recently moved. In 2025, the last complete year, Chinese organizations released 79 of the 164 frontier models in the sample and American organizations 68, shares of 48 and 41 percent \citep{epoch2026}. Usage runs in the same direction and further. Chinese open-weight models drew about 66 percent of activity on OpenRouter in July 2026, against 33 percent for United States developers \citep{openrouter2026}. Two cautions belong with these numbers. The 2026 release counts are a partial year, 23 models through July, too few to support a full-year share. The 2026 point in Figure \ref{fig:models} is provisional. The usage measure is a snapshot from a single routing platform, not a census of model use. Even with these caveats, the release and usage measures agree. The model layer is contested between two states, even as it remains competitive among firms.

\subsection{Chips and compute}
In the midstream band, concentration rises sharply and unevenly across the steps of making a chip (Table \ref{tab:chips}). The mildest is the design-tool layer, where three firms hold 74 percent and the index is 2{,}030. High-bandwidth memory, the fast memory stacked beside an AI accelerator, is a three-firm market shared by SK Hynix, Samsung, and Micron, with an index of 4{,}174 (Counterpoint Research). Higher still, concentration approaches the ceiling. Advanced packaging, the step that binds logic and memory into a working accelerator, is roughly 90 percent TSMC, an index of 8{,}100 (TrendForce). The extreme-ultraviolet scanners that print leading-edge logic come from ASML alone, placing that step at the ceiling of 10{,}000 \citep{khan2021,miller2022}. Several adjacent steps are reported only as a leading share rather than a full index, yet each is a near-chokepoint. Wafer-fabrication equipment is 70 percent in five firms, silicon wafers 90 percent in five firms, leading-edge logic fabrication 92 percent in Taiwan, and a single firm supplies about 86 percent of AI accelerators \citep{khan2021}. Compute is both the most demanded input and the most concentrated in its supply. Training compute has doubled roughly every six months over the deep-learning era \citep{sevilla2022}, while the hardware that provides it is held by a few firms \citep{sastry2024,vipra2023}.

\begin{table}[htbp]
\centering
\begin{threeparttable}
\caption{The chip and compute supply chain, step by step}
\label{tab:chips}
\small
\begin{tabular}{@{}lcc@{}}
\toprule
Step & Leading share & HHI \\
\midrule
Cloud infrastructure          & 63\% (top 3)     & 1{,}410 \\
Design tools (EDA)            & 74\% (top 3)     & 2{,}030 \\
High-bandwidth memory (HBM)   & $\approx$57\% (top firm) & 4{,}174 \\
Advanced packaging (CoWoS)    & 90\% (TSMC)      & 8{,}100 \\
EUV lithography               & 100\% (ASML)     & $\approx$10{,}000 \\
\addlinespace
AI accelerators               & 86\% (one firm)  & n.a. \\
Wafer-fab equipment           & 70\% (top 5)     & n.a. \\
Silicon wafers                & 90\% (top 5)     & n.a. \\
Leading-edge logic fab        & 92\% (Taiwan)    & n.a. \\
\bottomrule
\end{tabular}
\begin{tablenotes}[flushleft]\footnotesize
\item Here n.a.\ indicates that only a leading share is available, not a firm-level index. Sources, Synergy Research Group (cloud), TrendForce (design tools and packaging), Counterpoint Research (memory), CSET and \citet{khan2021} (lithography and logic fabrication), Yole (equipment), and analyst estimates (accelerators and wafers). All are 2025 market figures as of July 2026, and the market-tracker shares in particular move from quarter to quarter.
\end{tablenotes}
\end{threeparttable}
\end{table}

\subsection{Electricity and the Grid: An Upstream Problem}
\begin{figure}[htbp]
\centering
\begin{tikzpicture}
\begin{axis}[
  width=0.72\textwidth, height=0.42\textwidth,
  xlabel={Year}, ylabel={Net generation (TWh)},
  xtick={2017,2019,2021,2023,2025,2027},
  ymin=3800, ymax=4600,
  legend style={at={(0.02,0.98)}, anchor=north west, font=\footnotesize},
  grid=major, tick align=outside,
]
\addplot[thick,mark=*] coordinates {(2017,4034)(2018,4178)(2019,4128)(2020,4007)(2021,4110)(2022,4231)(2023,4183)(2024,4309)(2025,4430)};
\addlegendentry{Net generation (EIA-923)}
\addplot[thick,dashed,mark=square*] coordinates {(2025,4430)(2027,4478)};
\addlegendentry{Projected demand (STEO)}
\end{axis}
\end{tikzpicture}
\caption{United States electricity, annual net generation and projected demand. Generation rose from 4{,}007 terawatt-hours in 2020 to 4{,}430 in 2025. The Short-Term Energy Outlook projects 4{,}478 for 2027. The 2026 partial year is omitted. Source, U.S.\ Energy Information Administration \citep{eia2026}, as of July 2026.}
\label{fig:power}
\end{figure}
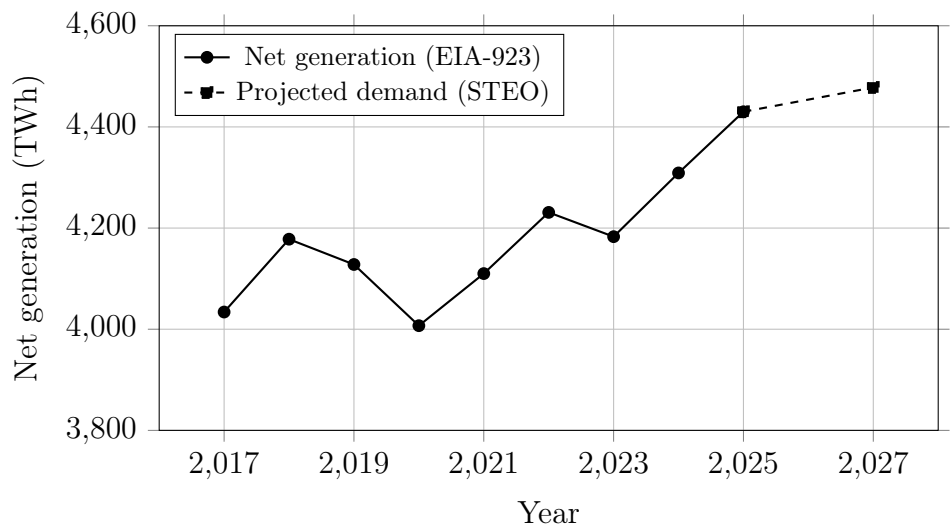

Electricity is the upstream-top layer, positioned between the chips it powers and the minerals it is generated from. It is measured differently from the other layers, because the binding constraint is a physical and administrative bottleneck rather than a market share. United States net generation, measured monthly through the EIA-923 survey, was 4{,}430 terawatt-hours in 2025, up from 4{,}007 in 2020. The Energy Information Administration's Short-Term Energy Outlook projects demand of 4{,}478 terawatt-hours in 2027 (Figure \ref{fig:power}) \citep{eia2026}. Against that envelope, data-center load is the fastest-growing component. It reached 4.4 percent of United States electricity in 2023 and is projected at 6.7 to 12 percent by 2028 \citep{shehabi2024}. Worldwide, data-center demand is projected to roughly double toward 945 terawatt-hours by 2030 \citep{ieaenergyai2025}.

The supply side cannot keep pace with this scenario. This is where the constraint binds. The typical new generator now waits about five years in the interconnection queue, up from under two years in 2008, with roughly 2{,}600 gigawatts of capacity stalled in the queue \citep{rand2024}. New transmission can take four to eight years to build \citep{ieaenergyai2025}. Power is therefore not concentrated in the sense the index captures, but it is becoming the layer that decides who can build new data centers, governed by physics and by the pace of approval rather than by any firm's market share.

Electricity is inseparable from the mineral and fuel layer beneath it. Whatever the source, generating power draws on that layer. Fossil generation burns extracted fuel. Natural gas, made abundant by the United States shale boom, is now the largest single source of American electricity. Coal remains a meaningful share. Both are mined rather than renewed. Nuclear generation runs on uranium that is mined, milled, and enriched. Enrichment is itself a concentrated step as discussed. Thorium, the leading alternative nuclear fuel, remains largely experimental. It too is mined, usually as a byproduct of rare-earth and heavy-mineral deposits. The renewable sources do not remove this dependence. They relocate it onto critical minerals. A solar panel is built from refined silicon, silver, and copper. A wind turbine is built from copper, steel, and, in direct-drive designs, rare-earth magnets. Grid-scale storage is built from lithium and graphite. Hydropower is the partial exception, heavy in concrete and steel rather than in critical minerals, though it is limited by geography and largely built out. The implication for this paper is direct. A shift from fossil fuels toward wind, solar, and storage does not lighten the grid's dependence on the upstream base. It moves that dependence from mined fuels toward refined critical minerals, which the paper finds among the most concentrated layers in the whole stack. Electricity and minerals therefore belong to a single upstream problem.

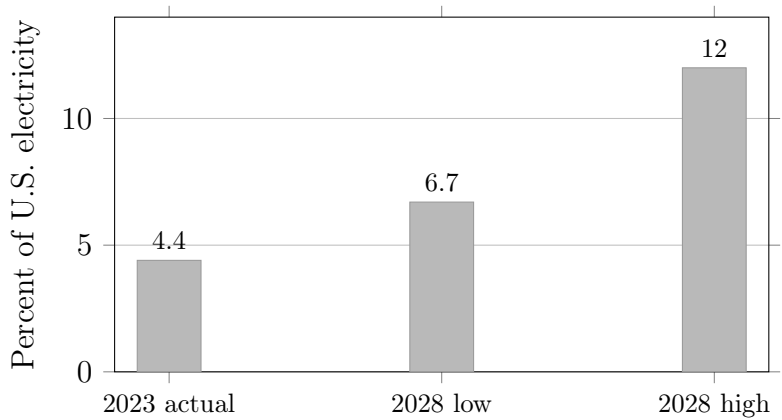
\begin{figure}[htbp]
\centering
\begin{tikzpicture}
\begin{axis}[
  ybar, width=0.62\textwidth, height=0.38\textwidth,
  xtick={1,2,3}, xticklabels={2023 actual,2028 low,2028 high},
  x tick label style={font=\footnotesize},
  ymin=0, ymax=14, ylabel={Percent of U.S. electricity},
  bar width=24pt, nodes near coords,
  every node near coord/.append style={font=\footnotesize},
  ymajorgrids, tick align=outside,
]
\addplot[fill=gray!55, draw=black!40] coordinates {(1,4.4) (2,6.7) (3,12)};
\end{axis}
\end{tikzpicture}
\caption{Data centers as a share of United States electricity. The 2028 figures are the published low and high projections. Source, Lawrence Berkeley National Laboratory \citep{shehabi2024}, as of July 2026.}
\label{fig:dcload}
\end{figure}

\subsection{Minerals and refining}
At the upstream base of the stack the unit of concentration is the country rather than the firm. The paper's least familiar finding is that concentration is sharper in refining than in mining. The pattern is not a single case but a regularity across the energy-chain minerals, the inputs that power, build, and cool the data centers (Table \ref{tab:minerals}). Cobalt is mined mostly in the Democratic Republic of the Congo, a production index of 5{,}562, but refined mostly in China, where the refining index is 6{,}241 and China's share is 79 percent. Rare earths are mined at an index of 5{,}054 with China at 69 percent and processed at 8{,}100 with China at 90 percent. Graphite is mined at 6{,}114 with China at 78 percent and processed at 8{,}649 with China at 93 percent. Even copper, whose mining is diffuse across Chile and Peru at an index of 1{,}206, refines at an index of 2{,}584. In each case the index rises as the material moves from the ground to the refinery. In each case the refinery is disproportionately in one country \citep{iea2025cmo,usgs2025,baskaran2025,nassar2020}. The compute-chain minerals, the inputs to the chips themselves, show the same country dominance already at the production stage. Silicon metal is 87 percent Chinese with an index of 7{,}590, and tungsten 79 percent Chinese at 6{,}242 \citep{usgs2025}.

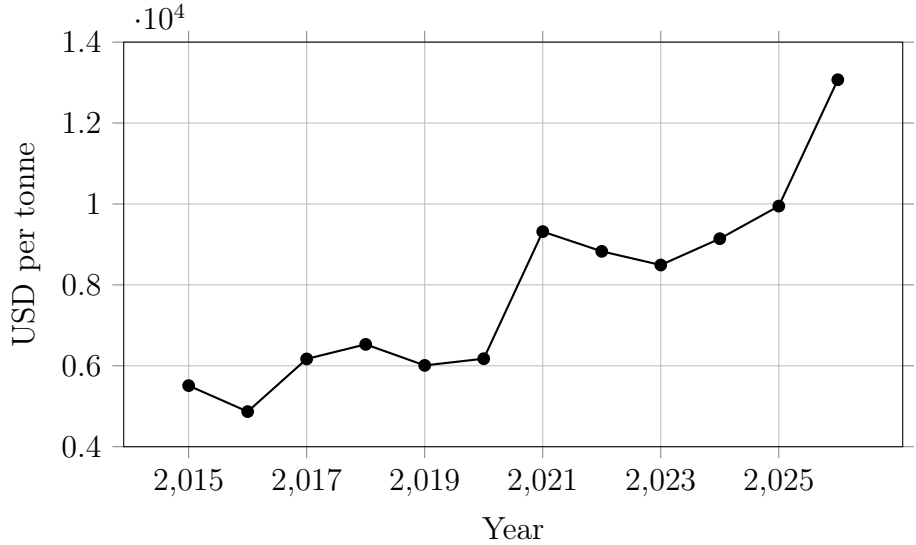
\begin{figure}[htbp]
\centering
\begin{tikzpicture}
\begin{axis}[
  width=0.72\textwidth, height=0.42\textwidth,
  xlabel={Year}, ylabel={USD per tonne},
  xtick={2015,2017,2019,2021,2023,2025},
  ymin=4000, ymax=14000,
  grid=major, tick align=outside,
]
\addplot[thick,mark=*] coordinates {(2015,5510)(2016,4868)(2017,6170)(2018,6530)(2019,6010)(2020,6175)(2021,9317)(2022,8829)(2023,8491)(2024,9142)(2025,9947)(2026,13070)};
\end{axis}
\end{tikzpicture}
\caption{World price of copper, annual average. The price more than doubled between 2020 and 2026. The 2026 value averages the first six months. Source, IMF via FRED \citep{fred2026}, as of July 2026.}
\label{fig:copperprice}
\end{figure}

Gallium is the limiting or exceptional case, because its production is itself a refining step. It is recovered as a byproduct of alumina rather than mined. China produces about 99 percent of the world's primary low-purity gallium, an index of 9{,}804 that rose from 9{,}401 in 2020 as the last non-Chinese output ceased \citep{usgs2025}. The leverage is not hypothetical. China placed gallium and germanium under export licensing in 2023 and banned their export to the United States in 2024, after which the price of gallium outside China rose sharply \citep{iea2025commentary,bown2025}. The most concentrated market in the entire AI stack is a mineral most readers cannot name, controlled by a single state that has already used that leverage. The broader price signal points the same way. Copper is a bellwether for electrification demand, of which data centers are one growing part. Its world price averaged about 6{,}175 dollars per tonne in 2020 and roughly 13{,}070 over the first half of 2026, more than doubling across the period (Figure \ref{fig:copperprice}) \citep{fred2026}. Table \ref{tab:countries} reports the country composition underlying each of these markets, which is what the indices summarize.

\begin{table}[htbp]
\centering
\begin{threeparttable}
\caption{Concentration rises from mining to refining}
\label{tab:minerals}
\small
\begin{tabular}{@{}lrrll@{}}
\toprule
Mineral & Prod.\ HHI '20 & Prod.\ HHI '25 & Lead producer '25 & Refining HHI \\
\midrule
\multicolumn{5}{@{}l}{\textit{Compute chain, inputs to the chips}}\\
Gallium       & 9{,}401 & 9{,}804 & China (99\%) & n.a. \\
Silicon metal & 6{,}297\tnote{*} & 7{,}590 & China (87\%) & n.a. \\
Tungsten      & 7{,}134 & 6{,}242 & China (79\%) & n.a. \\
\addlinespace
\multicolumn{5}{@{}l}{\textit{Energy chain, inputs to power and build data centers}}\\
Copper        & 1{,}325 & 1{,}206 & Chile, Peru  & 2{,}584 (China 48\%) \\
Rare earths   & 3{,}768 & 5{,}054 & China (69\%) & 8{,}100 (China 90\%) \\
Cobalt        & 4{,}865 & 5{,}562 & DRC (73\%)   & 6{,}241 (China 79\%) \\
Graphite      & 6{,}292 & 6{,}114 & China (78\%) & 8{,}649 (China 93\%) \\
\bottomrule
\end{tabular}
\begin{tablenotes}[flushleft]\footnotesize
\item Production indices are USGS Mineral Commodity Summaries estimates \citep{usgs2025}, 2025 edition as of July 2026. Refining and processing indices are for 2024, from the Cobalt Institute (cobalt), IEA (rare earths), and Benchmark Mineral Intelligence (graphite). Copper refining is from the USGS, 2025. Shares are the leading country.
\item[*] The silicon-metal series begins in 2022. The figure shown is the 2022 estimate.
\end{tablenotes}
\end{threeparttable}
\end{table}

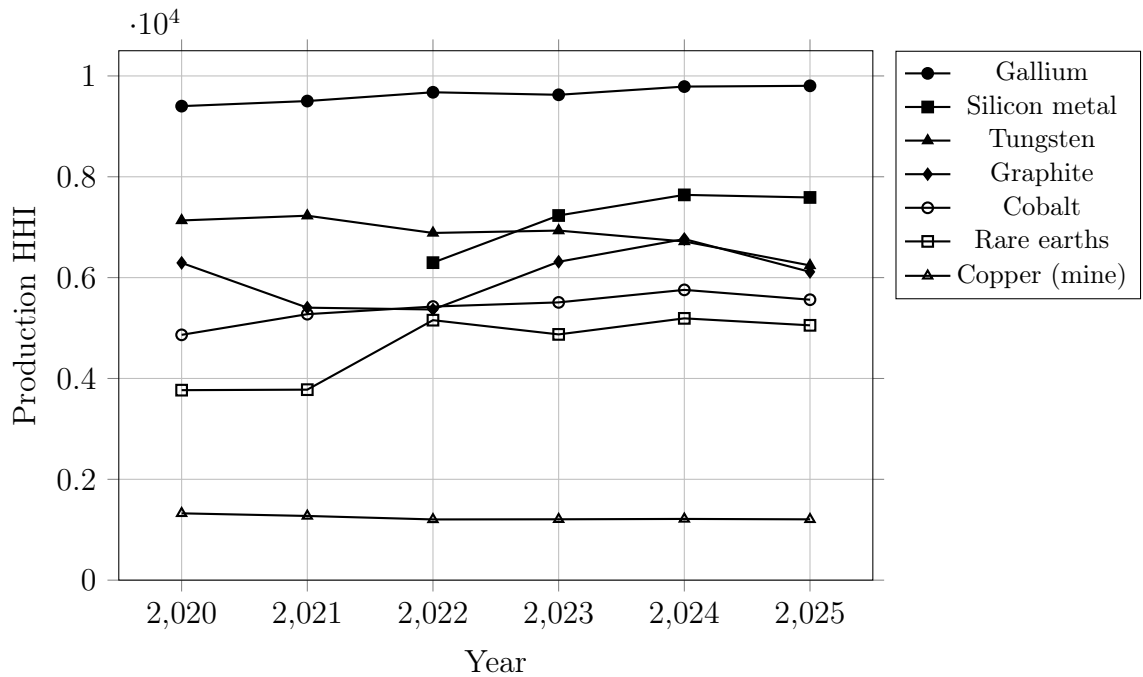
\begin{figure}[htbp]
\centering
\begin{tikzpicture}
\begin{axis}[
  width=0.70\textwidth, height=0.52\textwidth,
  xlabel={Year}, ylabel={Production HHI},
  xtick={2020,2021,2022,2023,2024,2025},
  ymin=0, ymax=10500,
  legend style={at={(1.03,1)}, anchor=north west, font=\footnotesize},
  grid=major, tick align=outside,
]
\addplot[thick,mark=*] coordinates {(2020,9401)(2021,9502)(2022,9676)(2023,9626)(2024,9789)(2025,9804)};
\addlegendentry{Gallium}
\addplot[thick,mark=square*] coordinates {(2022,6297)(2023,7232)(2024,7641)(2025,7590)};
\addlegendentry{Silicon metal}
\addplot[thick,mark=triangle*] coordinates {(2020,7134)(2021,7228)(2022,6886)(2023,6933)(2024,6717)(2025,6242)};
\addlegendentry{Tungsten}
\addplot[thick,mark=diamond*] coordinates {(2020,6292)(2021,5404)(2022,5368)(2023,6314)(2024,6765)(2025,6114)};
\addlegendentry{Graphite}
\addplot[thick,mark=o] coordinates {(2020,4865)(2021,5275)(2022,5426)(2023,5509)(2024,5757)(2025,5562)};
\addlegendentry{Cobalt}
\addplot[thick,mark=square] coordinates {(2020,3768)(2021,3778)(2022,5156)(2023,4874)(2024,5192)(2025,5054)};
\addlegendentry{Rare earths}
\addplot[thick,mark=triangle] coordinates {(2020,1325)(2021,1273)(2022,1204)(2023,1207)(2024,1214)(2025,1206)};
\addlegendentry{Copper (mine)}
\end{axis}
\end{tikzpicture}
\caption{Mineral production concentration, 2020 to 2025. Gallium remains near the ceiling throughout, silicon metal and rare earths rise over the period, and copper mining remains diffuse. The silicon series begins in 2022. Source, USGS Mineral Commodity Summaries \citep{usgs2025}.}
\label{fig:trajectories}
\end{figure}

\begin{figure}[htbp]
\centering
\begin{tikzpicture}
\begin{axis}[
  ybar, width=0.78\textwidth, height=0.45\textwidth,
  xtick={1,2,3,4}, xticklabels={Copper,Cobalt,Rare earths,Graphite},
  ymin=0, ymax=9800, ylabel={HHI},
  bar width=15pt,
  legend style={at={(0.02,0.98)}, anchor=north west, font=\footnotesize},
  nodes near coords, every node near coord/.append style={font=\scriptsize},
  ymajorgrids, tick align=outside,
]
\addplot[fill=gray!30, draw=black!40] coordinates {(1,1206) (2,5562) (3,5054) (4,6114)};
\addlegendentry{Mining or production}
\addplot[fill=gray!70, draw=black!40] coordinates {(1,2584) (2,6241) (3,8100) (4,8649)};
\addlegendentry{Refining or processing}
\end{axis}
\end{tikzpicture}
\caption{Concentration rises from mining to refining in all cases measured. Production indices are 2025. Refining and processing indices are 2024 for cobalt, rare earths, and graphite, and 2025 for copper. Sources as in Table \ref{tab:minerals}.}
\label{fig:refining}
\end{figure}
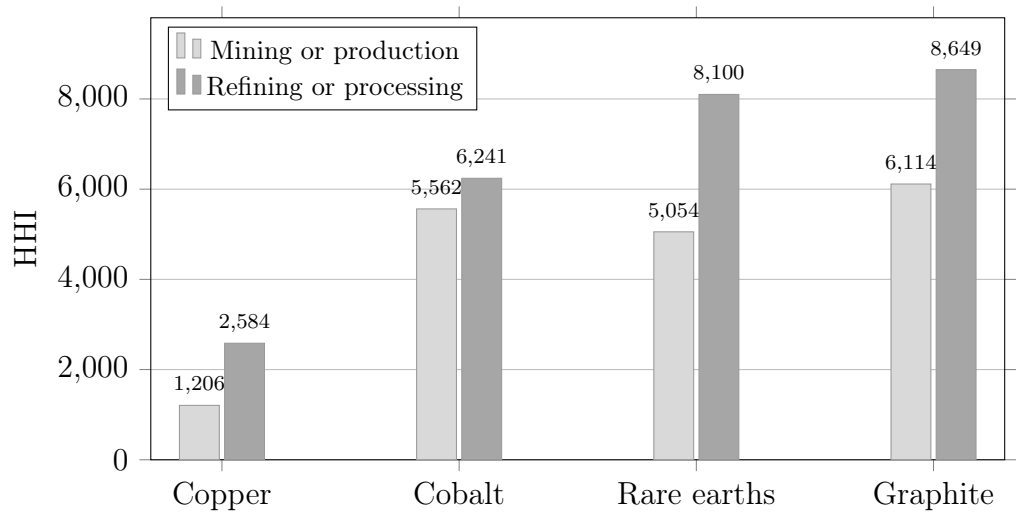

\begin{table}[htbp]
\centering
\begin{threeparttable}
\caption{Production by leading country, 2025}
\label{tab:countries}
\small
\begin{tabular}{@{}lrlll@{}}
\toprule
Mineral (unit) & World total & Leading producer & Second & Third \\
\midrule
Gallium (kg)          & 909{,}000     & China 99.0\%  & Russia 0.7\%         & Japan 0.3\% \\
Silicon metal (kt)    & 4{,}600       & China 87.0\%  & Brazil 3.9\%         & Norway 2.8\% \\
Tungsten (t)          & 85{,}100      & China 78.7\%  & Vietnam 3.5\%        & Kazakhstan 2.8\% \\
Rare earths (t)       & 390{,}000     & China 69.2\%  & United States 13.1\% & Australia 7.4\% \\
Cobalt (t)            & 314{,}600     & Congo 73.1\%  & Indonesia 14.0\%     & Russia 2.4\% \\
Graphite (t)          & 1{,}800{,}000 & China 77.8\%  & Madagascar 4.4\%     & Tanzania 4.2\% \\
Copper, mine (kt)     & 23{,}000      & Chile 23.0\%  & Congo 13.9\%         & Peru 11.7\% \\
Copper, refinery (kt) & 29{,}000      & China 48.3\%  & Congo 9.7\%          & Chile 5.9\% \\
\bottomrule
\end{tabular}
\begin{tablenotes}[flushleft]\footnotesize
\item 2025 estimates from the USGS Mineral Commodity Summaries \citep{usgs2025}, as of July 2026. The unit kt denotes thousand tonnes. World total is the figure these shares and the indices in Table \ref{tab:minerals} are measured against, taken as the larger of the published world total and the sum of listed producers. For several minerals the source reports a pooled residual of small producers. It is excluded from the ranking here and, following the convention in Section \ref{sec:framework}, is not squared when the index is computed, so each index is a lower bound.
\end{tablenotes}
\end{threeparttable}
\end{table}

\section{The Concentration Gradient and the Reach of Antitrust}\label{sec:gradient}
Table \ref{tab:hhi} and Figure \ref{fig:gradient} trace a clear gradient. Concentration is lowest at the downstream end of the stack, near 1{,}150 for model usage and 1{,}410 for cloud, where public and regulatory attention is highest. It rises steadily upstream, reaching 4{,}174 in memory, 8{,}100 in packaging, 9{,}804 in gallium, and the ceiling in lithography. This ordering is the paper's central finding. It is visible only when the layers are placed on one scale. Several of these layers are three to five times more concentrated than the 1{,}800 level at which United States agencies presume a merger to be anticompetitive.

The gradient has an uncomfortable corollary for competition policy. The layers that are most concentrated are the ones an antitrust authority can least reach. A merger review can shape the cloud and model layers, which are within domestic jurisdiction and near the guideline threshold. The recent turn of United States enforcement toward cloud-and-model partnerships is one such intervention \citep{ftc2025}. A merger review cannot subpoena a foreign state's refining capacity or compel a second firm into existence at the lithography layer. The tools that move the concentrated layers are export controls, industrial policy, and the diversification of supply, which is the language of economic statecraft rather than of merger control \citep{farrell2019,clayton2023,bown2025}. This is why the concentration that matters most for the resilience of the AI economy is also the concentration that the standard competition toolkit is least equipped to address. This distinction also changes how the model layer should be read. By firm, that layer is competitive, spread across many providers. By country, it is concentrated in the United States and China. Its firm-level competition is a question for antitrust. Its country-level concentration is a question for statecraft, the concern that dominates the upstream layers. The unit of power shifts as one moves up the stack, from the firm at the downstream end to the country at the upstream end. The model layer already carries both readings.

\section{Endowment, Capacity, and the Scope for Industrial Policy}\label{sec:geology}
\subsection{Geological endowment versus endogenous capacity}
A measurement of concentration does not tell a policymaker what to do. The right policy depends on whether the concentration is geological or chosen. If the reserves are diffuse but production is concentrated, the chokepoint was built rather than found. Industrial policy can move it. If the ore genuinely is not there, the levers are substitution and stockpiling instead. That distinction, between geology and choice, must be settled before any prescription. It is answerable, because the same source that reports production also reports reserves. This section places reserves on the same index as production and refining. For each material, it asks which of the two applies, and what that implies for policy.

\subsection{Reserves, resources, and prices}
The word reserve carries a specific meaning that the argument depends on. A reserve is not the quantity of a mineral in the earth's crust. It is the part of that quantity which can be extracted economically at current prices and with current technology. That makes it a smaller figure, and, importantly, a price-dependent one \citep{usgs2026}. Two consequences follow. Both strengthen the case that concentration is chosen rather than fixed.

The first consequence concerns gallium and silicon. The USGS reports no reserves for either. This reflects abundance rather than scarcity. Gallium has no deposit of its own. It occurs only as a trace in bauxite and is recovered as a byproduct of alumina refining rather than mined. The gallium held in world bauxite, estimated at more than a million tonnes, is spread wherever bauxite is found. Silicon comes from quartz, which the USGS describes as abundant and adequate to supply world needs for many decades. For both, a reserve figure of zero does not mean the material is scarce. It means scarcity is not the constraint at all. With no scarcity to invoke, their concentration must come from built capacity rather than from geology.

The second consequence follows from the price definition. A reserve is therefore not a fixed geological quantity. At a given price it is a definite number. That number rises when the price rises. A restriction that pushes the world price up can turn deposits that were uneconomic into reserves.

Whether this correction works depends on how concentrated the reserves are. That concentration fixes the market structure at the reserve stage. It is what the reserves index measures. When the ore is spread across many countries, the reserve market is competitive. No single holder can influence the price. If one restricts supply, the others expand as price-takers. In the long run the price returns to the competitive level. Copper, with a reserves index of 745, is this case. When the ore is held by only a few countries, the reserve market is an oligopoly. Cobalt, tungsten, and rare earths each have one country holding close to half the reserves, with a reserves index between roughly 2{,}800 and 3{,}400. If the leading holder restricts supply, the others can expand. As oligopolists, though, they can hold the price above the competitive level and earn a margin over cost \citep{askercollardwexlerdeloecker2019}. The size of that margin depends on their conduct. Under non-cooperative competition it stays small. Under collusion, in effect a cartel, it approaches the monopoly price.

Two limits qualify this. The first is that a rising price expands reserves faster than it expands output. Producing from a reserve is slow. Output from existing operations responds little to price. Adding new capacity takes years \citep{andersonkelloggsalant2018}.\footnote{The empirical studies cited across this subsection, including the market-power estimate above, are for petroleum rather than for the hard-rock critical minerals such as cobalt and the rare earths at issue in this paper. A new mine comes online far more slowly than a new well, on the order of a decade or more from discovery to first production and four to five years for construction alone at a known deposit \citep{iea2021minerals}, against the weeks to a few months required to drill a shale well and bring it into production \citep{eia2019drilling}. Current mine production and processing, the stages that govern the short-run supply response, are also more concentrated than in oil, with China accounting for about 70 percent of rare-earth mining and roughly 90 percent of rare-earth refining and the Democratic Republic of the Congo for about 73 percent of mined cobalt \citep{usgs2026, iearee2026}, shares that stand well above the 32 to 40 percent of world crude output that the OPEC countries supply \citep{eia2023opec}. On these two dimensions, the slow addition of capacity and the concentration of supply, the petroleum estimates understate rather than overstate the rigidity of mineral supply and the market power available to its leading producers, a conservative direction of bias for the present case.} Opening a mine requires permits, access to land, and passage through environmental and social review. On public land it also depends on the leasing and auction institutions that decide how much is developed \citep{kim2026}. Where the ore lies under tribal land or communities that resist displacement, opposition alone can stall a project for years. Building the refinery the ore then needs is slower still. Refining is the most concentrated stage. No firm will build against a dominant incumbent that can undercut it, one of the market failures set out below. Mining and refining are both chokepoints. The correction reaches them over years rather than months.

The second limit is the mineral's share of the cost of the product made from it, which by equation~\eqref{eq:passthrough} sets how much of a price rise reaches that product. For compute hardware the share is small, because the value of a chip lies in its design and fabrication rather than in its raw materials. Gallium is a negligible part of a finished chip's cost. Even a large rise in its price barely changes that cost. The threat from a hardware mineral is therefore its availability rather than its price. A supply cutoff can halt production even when the material is cheap. The power layer is a partial exception. As an operating cost, electricity is the largest single category of the cost of running a model. A rise in its price therefore feeds through to that cost. Measured against the full cost of ownership, which also counts the hardware capital, that capital dominates. Electricity is a small share of that total \citep{epochtco2026}. Across the whole stack, then, the threat runs through access to the input rather than through its cost. At the power layer that access constraint is the physical limit on adding grid capacity examined earlier.

\subsection{The distribution of reserves, production, and refining}
\begin{figure}[htbp]
\centering
\begin{tikzpicture}
\begin{axis}[
  ybar, width=0.85\textwidth, height=0.45\textwidth,
  symbolic x coords={Copper,Graphite,Cobalt,Tungsten,Rare earths},
  xtick=data, x tick label style={font=\footnotesize},
  ymin=0, ymax=9600, ylabel={HHI},
  bar width=9pt,
  legend style={at={(0.02,0.98)},anchor=north west,font=\footnotesize},
  ymajorgrids, tick align=outside,
]
\addplot[fill=gray!20,draw=black!40] coordinates {(Copper,745)(Graphite,1833)(Cobalt,2815)(Tungsten,3064)(Rare earths,3394)};
\addlegendentry{Reserves (ground)}
\addplot[fill=gray!50,draw=black!40] coordinates {(Copper,1206)(Graphite,6114)(Cobalt,5562)(Tungsten,6242)(Rare earths,5054)};
\addlegendentry{Production (mine)}
\addplot[fill=gray!80,draw=black!40] coordinates {(Copper,2584)(Graphite,8649)(Cobalt,6241)(Rare earths,8100)};
\addlegendentry{Refining}
\draw[dashed] (axis cs:Copper,1800) -- (axis cs:Rare earths,1800);
\node[anchor=south east,font=\scriptsize] at (axis cs:Rare earths,1800) {1{,}800 line};
\end{axis}
\end{tikzpicture}
\caption{Concentration rises from the ground to the refinery. For all minerals with reserves, the index is lowest for reserves, higher for production, and higher still for refining. It crosses the 1{,}800 highly-concentrated line early. Tungsten refining is not separately measured. Sources, USGS \citep{usgs2026} and Table \ref{tab:minerals}.}
\label{fig:threestage}
\end{figure}
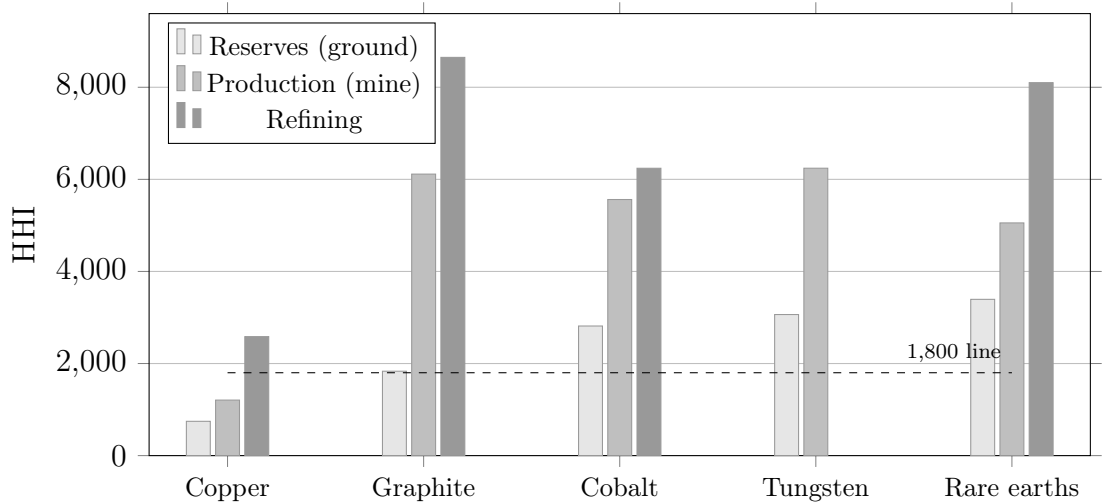

Table \ref{tab:reserves} places the reserves on the index and sets them beside production and refining. Figure \ref{fig:threestage} plots the three stages together. The result is uniform, and it is the section's central fact. In all minerals for which reserves are reported, concentration is lowest in the ground, higher at the mine, and higher still at the refinery. The ordering does not reverse in a single case.

The ore is far more widely distributed than the mines, and the mines far more than the refineries. China holds 32 percent of graphite reserves but processes 93 percent of the graphite. It holds essentially no cobalt ore, 1.3 percent of world reserves, yet refines 79 percent of the world's cobalt. Brazil holds a quarter of world rare-earth reserves and 74 million tonnes of graphite, the second-largest reserve of each, and produces almost none of either. The chokepoint was built, not found. That makes it a policy variable rather than a geological fact.

The distribution repays reading in detail, because it decides the policy for each material. For copper the reserves are genuinely diffuse, with Chile the largest holder at 18 percent and no country above a fifth. The concentration that exists is mild and appears only at the refinery. For graphite, China holds barely a third of reserves against Brazil's quarter and a further fifth spread across Madagascar, Mozambique, and Tanzania. The ore is not the chokepoint at all. For rare earths, China holds half the reserves, but Brazil, Australia, Vietnam, the United States, Greenland, and Canada hold the rest. The United States already mines 13 percent of world output at a single site. For tungsten, China holds half the reserves, with a substantial allied tail in Australia, Spain, Austria, and Portugal. Cobalt is the one exception, the material whose ore is genuinely concentrated, with the Democratic Republic of the Congo holding half the reserves and mining 73 percent of world output. Even there, the refining concentration is a separate layer built on top of the geological one.

\begin{table}[htbp]
\centering
\begin{threeparttable}
\caption{Reserves, production, and refining on one index}
\label{tab:reserves}
\small
\begin{tabular}{@{}lllrrr@{}}
\toprule
Mineral & World & Leading reserve holders & Res. & Prod. & Refin. \\
 & reserves & (\% of world reserves) & HHI & HHI & HHI \\
\midrule
Copper      & 980 Mt  & Chile 18, Australia 10, Peru 9    & 745    & 1{,}206 & 2{,}584 \\
Graphite    & 310 Mt  & China 32, Brazil 24, Madagascar 9 & 1{,}833 & 6{,}114 & 8{,}649 \\
Cobalt      & 12 Mt   & Congo 50, Australia 14, Russia 7  & 2{,}815 & 5{,}562 & 6{,}241 \\
Tungsten    & 4.7 Mt  & China 53, Australia 12, Russia 9  & 3{,}064 & 6{,}242 & n.a. \\
Rare earths & 85 Mt   & China 52, Brazil 25, Australia 7  & 3{,}394 & 5{,}054 & 8{,}100 \\
\addlinespace
Gallium       & \textit{none} & byproduct, bauxite resources $>$1 Mt & n.a. & 9{,}804 & n.a. \\
Silicon metal & \textit{none} & smelted from ubiquitous quartz       & n.a. & 7{,}590 & n.a. \\
\bottomrule
\end{tabular}
\begin{tablenotes}[flushleft]\footnotesize
\item Reserves are 2025 estimates from the USGS Mineral Commodity Summaries \citep{usgs2026}, as of July 2026. Mt denotes million tonnes of contained metal or mineral. Reserve holders are the three largest, in percent of world reserves. Reserves and production indices are the author's calculation under the pooled-residual convention of Section \ref{sec:framework}, so each is a lower bound. Refining indices are as in Table \ref{tab:minerals}. Gallium and silicon carry no reserves because gallium is a byproduct with no deposit of its own and silicon is smelted from ubiquitous quartzite, discussed below.
\end{tablenotes}
\end{threeparttable}
\end{table}

\subsection{Expanding primary supply}
Where the concentration is a built capacity rather than a natural endowment, an alternative can be created. There are four ways to do it, which are not interchangeable. The distinction between them is the practical content of the policy, because each fits a different material and carries a different cost.

The first is to \textit{mine}, to extract from a domestic or allied ore body. This is available only where reserves exist, which the previous subsection has shown is most of the materials. It already happens at the margin, as with United States and Australian rare-earth mining. But the reserves must also clear the domestic land and permitting regime before they can be worked, a constraint taken up below. The second is to \textit{refine}, to build the refining capacity that turns ore or an intermediate into the usable input. This is the step that is most concentrated and the one that geology never constrains. It is the binding response for graphite, rare earths, and cobalt, where the ore or the mined supply is available but the processing is not. The Congo-to-China pattern, in which one country mines and another refines, is precisely the pattern a refining build-out would break. The third is to \textit{recover}, to extract a byproduct from the processing stream of another material, which applies only to gallium and germanium. This lever is unusual in requiring neither a new mine nor a new refinery, only a recovery circuit added to plants that already run. The West refines about a third of the world's alumina while recovering almost none of the gallium dissolved in it. A peer-reviewed assessment puts latent supply from existing bauxite and zinc streams at five times current production, limited by installed capacity rather than by geology \citep{frenzel2016,usgs2026}. The fourth is to \textit{manufacture}, to synthesize the functional material from an abundant feedstock. Synthetic graphite is made from petroleum coke. Silicon is smelted from quartz. Here the lesson is a caution. Manufacturing does not eliminate the concentration. It relocates it to whatever input is scarce and immobile, which for both is cheap electricity. China holds a higher share of battery-grade synthetic graphite, 91 to 94 percent, than of mined natural graphite. Switching to the manufactured material raises its dominance rather than escaping it \citep{iea2026cmo}.

\subsection{Substitution, recycling, and diversification}
Three further responses reduce the dependence rather than expand the supply, and they sort the materials into two regimes.

Cobalt and copper have real margins that are already being exercised. Lithium iron phosphate cathodes contain no cobalt and accounted for more than 55 percent of the electric-vehicle battery market in 2025, up from about 10 percent in 2020 \citep{ieaev2026}. Cobalt recovered from scrap equals a quarter of United States consumption \citep{usgs2026}. Recycled copper supplies about a third of world use \citep{icsg2025}. Aluminium substitutes for copper in overhead lines and heat exchangers. Rare earths and tungsten have no such margin. There the concentrated foreign supply is a capability problem rather than a price problem. The rare-earth-free ferrite magnet delivers roughly a tenth of the magnetic strength of a neodymium magnet and is not a drop-in replacement at the performance frontier \citep{iearee2026,doe2022}. End-of-life magnet collection falls below 5 percent. And tungsten's substitutes, in the USGS's own words, reduce rather than replace the amount of tungsten used.

One discipline runs through all of this and bears on any model of adjustment. The substitution that actually occurs is thrifting within a technology rather than switching away from the element, a movement along the intensive margin rather than the extensive one. Rare-earth use per unit is nearly 30 percent below its 2010 level. The United States Department of Energy estimates that a doubling of price would cut the heavy-rare-earth content of certain magnets by about 43 percent \citep{iearee2026,doe2022}. The cleanest evidence is the copper-to-aluminium price ratio, long treated as crossing a substitution threshold at three to one. That ratio has remained above three and a half to one for most of the past five years without producing the switch \citep{iea2026cmo}. Treating substitution as a binary would overstate the economy's capacity to adjust, because the measured elasticity lies in the intensity of use and not in the choice of material.

The third response, diversification, buys the same material from a friendlier holder of reserves or capacity. The geological table names the candidates. Brazil for graphite and rare earths, Australia for cobalt and tungsten, and the Western alumina refiners for gallium are not hypothetical alternatives but existing holdings that current trade patterns bypass. Table \ref{tab:policy} sets out all seven responses, material by material, so that none is left implicit.

\begin{table}[htbp]
\centering
\begin{threeparttable}
\caption{The response ledger, material by material}
\label{tab:policy}
\footnotesize
\begin{tabular}{@{}lp{3.4cm}p{3.0cm}p{1.9cm}p{2.5cm}@{}}
\toprule
Material & Make more (primary supply) & Substitute & Recycle & Diversify to \\
\midrule
Gallium       & Recover from alumina and zinc streams, no mine needed & None material & New scrap only & Western alumina refiners \\
Silicon metal & Smelt from abundant quartz, binding input is power & None material & n.a. & Any low-power-cost region \\
Graphite      & Mine (reserves diffuse) or manufacture synthetic & Synthetic for natural & Limited & Brazil, Africa \\
Rare earths   & Mine allied reserves, then build refining & Ferrite about one tenth the strength, no drop-in & Under 5\% end-of-life & Brazil, Australia, Vietnam \\
Cobalt        & Refine domestically, the ore is Congo's & LFP cathodes use none, over half of EV batteries & 25\% of U.S.\ use & Australia, Indonesia, Canada \\
Tungsten      & Mine some allied reserves, then refine & Substitutes reduce, not replace & About 25\% scrap & Australia, Vietnam, Iberia \\
Copper        & Mine (diffuse) and refine & Aluminium at about 60\% conductivity & About one third of use & Already diffuse \\
\bottomrule
\end{tabular}
\begin{tablenotes}[flushleft]\footnotesize
\item Sources as cited in Section \ref{sec:geology}, principally \citet{usgs2026}, \citet{iea2026cmo}, \citet{ieaev2026}, \citet{iearee2026}, \citet{iearecycling2024}, \citet{doe2022}, \citet{icsg2025}, and \citet{frenzel2016}. Figures are as of July 2026.
\end{tablenotes}
\end{threeparttable}
\end{table}

\subsection{Market failures and the rationale for intervention}
If a refinery is only a capital stock, a natural objection is that the private market should already have rebuilt it somewhere cheaper or safer. The fact that it has not is a puzzle any policy claim must answer. The industrial-policy literature supplies the answer. It is a set of market failures rather than an appeal to security alone.

The first is a coordination failure of the kind formalized in the big-push tradition \citep{murphyshleifervishny1989,klinemoretti2014}. A new refinery is viable only if feedstock and offtake are both assured. No single firm will build it against a dominant incumbent that can cut its price below cost until the entrant fails. That is not a hypothetical. Germany, Hungary, and Kazakhstan each abandoned gallium production when cheaper Chinese material appeared, and Western silicon smelters closed when power prices rose, in each case an individually rational exit that left the market to one country. The second failure is learning-by-doing, the dynamic economy of scale in which unit cost falls with cumulative output \citep{arrow1962}. Processing know-how of this kind is what two decades of Chinese investment in refining has accumulated. An entrant therefore meets not a level field but an incumbent far down its learning curve. The theory of dynamic comparative advantage shows that temporary support can reverse the long-run outcome in the entrant's favor \citep{krugman1987,greenwaldstiglitz2006,hausmannrodrik2003}. The third is the strategic leverage a dominant supplier holds, the chokepoint and holdup power modeled in the geoeconomics literature \citep{clayton2023,farrell2019}, which private firms do not internalize and which raises the social return to a domestic alternative above the private one.

That these failures can be corrected, and that targeted support can build a comparative advantage rather than merely subsidize a losing one, is the central lesson of a decade of credible industrial-policy evaluation \citep{juhaszlanerodrik2024}. South Korea's heavy-and-chemical drive raised the output and the dynamic comparative advantage of its targeted sectors, with gains that persisted for decades after the subsidies ended \citep{lane2025,choilevchenko2025}. Temporary protection during the Napoleonic blockade durably raised French mechanized industry \citep{juhasz2018}, war reparations pushed Finnish regions into heavy industry with lasting income and mobility gains \citep{mitrunen2025}, and the Tennessee Valley Authority's big push left manufacturing employment permanently higher \citep{klinemoretti2014}. The reassessment is not unconditional. The best current estimates put the welfare gains from optimal intervention at real but bounded magnitudes \citep{bartelme2025}. The evidence favors targeted and accountable forms of policy over broad price distortions \citep{harrisonrodriguezclare2010,rodrik2008}. A paper that has located the precise layers where concentration binds is well placed to supply the targeting that discipline requires.

Two instruments follow from the diagnosis. Where the problem is coordination and holdup, a guaranteed offtake or a price floor lets a refinery be financed against the threat of a price war, which undoes the market failure rather than merely subsidizing production. For the residual cases of cobalt ore, rare-earth processing, and tungsten, supply cannot be rebuilt on any short horizon. The appropriate tool is a strategic stockpile. The economics of a stockpile as a buffer against scarcity and price spikes is long established \citep{wrightwilliams1982,deatonlaroque1992}. A further constraint underlies all of this, one that the reserve figures conceal. The right to extract on public land is allocated by leasing and auction institutions. The design of those institutions governs how much is actually developed \citep{kim2026}. Permitting and leasing reform is a lever on the mining response in the same way that an offtake guarantee is a lever on the refining response.

\subsection{Matching instruments to materials}
Table \ref{tab:policy} collects the diagnosis. Three groups emerge, each calling for a different instrument.

Where there is no reserve constraint at all, gallium and silicon, the concentration is purely a matter of who built and operated capacity. The corrective is investment in recovery circuits and competitive industrial electricity, not mining. Gallium can be recovered from alumina streams the West already refines. The wave of recovery projects announced since the 2023 export controls, in Australia, Canada, Greece, Kazakhstan, Korea, and the United States, requires no new mine \citep{usgs2026}. Where reserves are widely held but processing is not, as with graphite, rare earths, and copper, the corrective is refining capacity together with the permitting and offtake certainty that lets it be financed. The latent alternatives are identifiable, with Brazil alone holding roughly a quarter of world reserves of both graphite and rare earths. Where the endowment itself is concentrated, cobalt ore in the Democratic Republic of the Congo, mining policy cannot resolve the problem. The workable levers are the substitution already under way through cobalt-free chemistries, together with recycling and stockpiles.

The uncomfortable residual is the pair for which neither geology nor substitution offers relief. Rare-earth processing and tungsten combine a concentrated refining step with substitutes that degrade performance by roughly an order of magnitude. These are the two places among the mineral layers where a supply interruption cannot be engineered around on any short horizon. They are where a measurement exercise alone, without the reserves and substitution data laid beside it, would have failed to direct policy attention. That is the reach of industrial policy in a sentence. It is wide for gallium and silicon, real but capital-intensive for the refined minerals, and narrow for cobalt ore, rare-earth processing, and tungsten. Knowing which is which is the payoff of putting endowment, capacity, and substitution on one page. It is the difference between a paper that measures a problem and one that says what to do about it.

\section{Limitations and Extensions}\label{sec:limits}
The paper measures the concentration of production and processing, not firm-level market power at all steps. The two can diverge. A high share need not imply the ability to raise price where entry is easy or the buyer is itself powerful \citep{shapiro2019}. Several layers are reported at the level of the country rather than the firm, where the relevant unit of power is a state and the appropriate lens is economic statecraft rather than antitrust. The figures also differ in provenance, which the confidence tiers record. Frontier-model usage is a proxy drawn from one routing platform and is a snapshot that moves from month to month. The cloud figure is a revenue share standing in for installed compute. Several chip steps are analyst or trade-tracker estimates rather than audited counts. The mineral processing shares come from industry bodies at annual frequency. None of these carries the authority of the primary official statistic that the USGS provides for mineral production. The paper marks them accordingly rather than presenting a false uniformity of precision.

The market-definition caution from the concentration literature applies here in a useful way \citep{rossihansberg2021,benkard2026}. Those papers show that a concentration measure can be inflated by drawing the market too narrowly or deflated by drawing it too broadly. The chokepoints measured here are defined at the narrow physical step at which substitution actually binds, a single EUV scanner design, a single packaging line, a single refined feedstock, so the high indices reflect a property of the technology rather than an artifact of aggregation. Where a layer could reasonably be drawn more broadly, as with cloud, the paper reports the broader and therefore lower figure, which biases the gradient toward understatement rather than exaggeration.

The extensions the data most naturally support would deepen the picture without altering its shape. Several compute steps, the accelerators, the wafer and equipment markets, and leading-edge fabrication, are now reported only as a leading share. Firm-level indices for them would place all the chip layers on the index rather than a subset of them. A bounded range for the AI-driven component of copper and electricity demand would separate the growth attributable to data centers from the rest. And a formal treatment of the interconnection queue as a capacity constraint would turn the qualitative grid bottleneck into a measured one. Each of these is on the public roadmap of the dashboard. Each would sharpen rather than overturn the gradient documented here.

\section{Conclusion}\label{sec:conclusion}
The competition debate in artificial intelligence has trained its attention on the model layer, which the evidence here shows to be the least concentrated and most contestable part of the stack. Placed on a single index, concentration rises the further upstream one looks, from about 1{,}150 at model usage and 1{,}410 at cloud to 4{,}174 in memory, 8{,}100 in packaging, 9{,}804 in gallium, and the ceiling at lithography. The physical inputs at the upstream end of the stack are three to five times more concentrated than the level at which United States regulators presume a merger to be a problem. They are the inputs a merger review cannot reach.

The model layer, read two ways, is the whole argument in miniature. By firm it is competitive, with an index near 500. By country it is concentrated and becoming more so, with an index above 5{,}000. Its usage has already tilted toward one of those countries. As one moves upstream the firm-level question fades and the country-level question sharpens, until at the upstream base the relevant actor is a single state with a near-monopoly on a refined input and a demonstrated willingness to restrict its export. The instruments that govern that kind of concentration are export controls, industrial policy, and the patient diversification of supply, not the merger guidelines.

The methodological point is as important as the substantive one. The layers of the AI stack have been studied in separate literatures, in incompatible units, often from proprietary data. Placing them on one index with public and reproducible provenance is what makes the gradient visible and what lets any figure in this paper be checked against its source. If the goal is to understand where market power in artificial intelligence actually resides, the first requirement is to measure the whole stack on a common scale. The result of doing so is plain. The competitive risk is greatest upstream, where public and regulatory scrutiny is least.

\newpage
\bibliography{refs}

\end{document}